\documentstyle[12pt,bezier]{article}
\def\beq{\begin{equation}}
\def\eeq{\end{equation}}

\def\ap#1#2#3 {Ann. Phys. (NY) {\bf#1} (19#2) #3}
\def\apj#1#2#3 {Astrophys. J. {\bf#1} (19#2) #3}
\def\apjl#1#2#3 {Astrophys. J. Lett. {\bf#1} (19#2) #3}
\def\app#1#2#3 {Acta. Phys. Pol. {\bf#1} (19#2) #3}
\def\ar#1#2#3 {Ann. Rev. Nucl. Part. Sci. {\bf#1} (19#2) #3}
\def\cpc#1#2#3 {Computer Phys. Comm. {\bf#1} (19#2) #3}
\def\err#1#2#3 {{\it Erratum} {\bf#1} (19#2) #3}
\def\ib#1#2#3 {{\it ibid.} {\bf#1} (19#2) #3}
\def\jmp#1#2#3 {J. Math. Phys. {\bf#1} (19#2) #3}
\def\ijmp#1#2#3 {Int. J. Mod. Phys. {\bf#1} (19#2) #3}
\def\jetp#1#2#3 {JETP Lett. {\bf#1} (19#2) #3}
\def\jpg#1#2#3 {J. Phys. G. {\bf#1} (19#2) #3}
\def\mpl#1#2#3 {Mod. Phys. Lett. {\bf#1} (19#2) #3}
\def\nat#1#2#3 {Nature (London) {\bf#1} (19#2) #3}
\def\nc#1#2#3 {Nuovo Cim. {\bf#1} (19#2) #3}
\def\nim#1#2#3 {Nucl. Instr. Meth. {\bf#1} (19#2) #3}
\def\np#1#2#3 {Nucl. Phys. {\bf#1} (19#2) #3}
\def\pcps#1#2#3 {Proc. Cam. Phil. Soc. {\bf#1} (#2) #3}
\def\pl#1#2#3 {Phys. Lett. {\bf#1} (19#2) #3}
\def\prep#1#2#3 {Phys. Rep. {\bf#1} (19#2) #3}
\def\prev#1#2#3 {Phys. Rev. {\bf#1} (19#2) #3}
\def\prl#1#2#3 {Phys. Rev. Lett. {\bf#1} (19#2) #3}
\def\prs#1#2#3 {Proc. Roy. Soc. {\bf#1} (19#2) #3}
\def\ptp#1#2#3 {Prog. Th. Phys. {\bf#1} (19#2) #3}
\def\ps#1#2#3 {Physica Scripta {\bf#1} (19#2) #3}
\def\rmp#1#2#3 {Rev. Mod. Phys. {\bf#1} (19#2) #3}
\def\rpp#1#2#3 {Rep. Prog. Phys. {\bf#1} (19#2) #3}
\def\sjnp#1#2#3 {Sov. J. Nucl. Phys. {\bf#1} (19#2) #3}
\def\spj#1#2#3 {Sov. Phys. JEPT {\bf#1} (19#2) #3}
\def\spu#1#2#3 {Sov. Phys.-Usp. {\bf#1} (19#2) #3}
\def\zp#1#2#3 {Zeit. Phys. {\bf#1} (19#2) #3}
\begin{document}
\begin{titlepage}
\begin{center}
{\Large \bf Theoretical Physics Institute \\
University of Minnesota \\}  \end{center}
\vspace{0.3in}
\begin{flushright}
TPI-MINN-94/35-T \\
UMN-TH-1316-94 \\
September 1994
\end{flushright}
\vspace{0.4in}
\begin{center}
{\Large \bf QCD radiative enhancement of the decay $b \to  c\,{\overline
 c}\, s$. \\}
\vspace{0.2in}
{\bf M.B. Voloshin  \\ }
Theoretical Physics
Institute, University of Minnesota \\ Minneapolis, MN 55455 \\ and \\
Institute of Theoretical and Experimental Physics  \\
                         Moscow, 117259 \\
\vspace{0.3in}
{\bf   Abstract  \\ }
\end{center}
A substantial enhancement is found of the $b \to  c\,{\overline c}\, s$
decay rate due to the QCD interaction within  the  $\overline c
s$ pair, which enhancement may amount to 30\%, or more. Some general
features of calculation of the QCD radiative corrections in two first orders
are discussed.

\end{titlepage}

The present theoretical understanding of $B$ decays conspicuously runs into
a problem with explaining the experimental observation of rather low
semileptonic branching ratio$^{\cite{slexp}}$, which
requires$^{\cite{baffling}}$ an enhancement of the non-leptonic decay rate
by as much as 20\% - 30\% over the existing theoretical predictions.
Therefore a more thorough theoretical study of the $B$ decays is imperative.
In this note the decay $b \to c\,{\overline c}\, s$ is discussed, whose rate
is suppressed by the presence of two massive charmed quarks as compared to
the dominant non-leptonic decay $b \to  c\,{\overline u}\, d$. An
enhancement of the decay $b \to c\,{\overline c}\, s$ would somewhat relax
the problem of low semileptonic branching ratio. However, such enhancement
would also worsen another possible problem, which is perhaps hinted at by
the experiment$^{\cite{cyield}}$, that is the problem of a low average charm
yield per $B$ decay\footnote {Clearly, it is impossible to solve
simultaneously both these problems by an enhancement of the decay $b \to
c\,{\overline c}\, s\,^{\cite{fwd}}$}.
Nevertheless, within the present uncertainty of the
measured average charm yield in $B$ decays the data can still accommodate a
substantial enhancement of the decay $b \to c\,{\overline c}\, s$.  In any
case, it is important to have the potentially essential effects calculated.
Here it is found that the first order QCD correction to the total rate of
the decay $b \to c\,{\overline c}\, s$ due to the interaction within the
quark pair ${\overline c}\, s$ has relative magnitude $\delta_{\overline c
s} \, \alpha_s/\pi$ with an unusually large coefficient:  $\delta_{\overline
c s} = 4.46$ for $m_c/m_b=0.3$, unlike the case of the decay $b \to
c\,{\overline u}\, d$, where the similar factor due to the ${\overline u}\,
d$ pair is equal to one.

The QCD radiative effects in the
decay rate of the $b$ quark are usually analyzed within the leading
log approximation in $\ln (m_W/m_b)$ or in the next-to-leading log
approximation (see e.g. \cite{ap,bbbg}). However, since, $\ln (m_W/m_b)$ is
not really a sufficiently large parameter, the non-logarithmic terms may be
quite essential, and it might be more reasonable to rely instead on complete
calculation of the QCD radiative corrections in the first and second orders
in $\alpha_s$. The error in the logarithmic terms, induced by such
truncation of the series, is then not more than about 5\% (in the total
rate), which is not larger than other uncertainties in the calculation, in
particular not larger than the non-logarithmic terms.

Both the perturbative and non-perturbative effects in the inclusive decay
rates of $B$ hadrons are conveniently calculated by
representing$^{\cite{sv}}$ the decay rate through the absorptive part of the
$B$
self-energy, arising in the second order in the weak interaction. For the
first-order QCD corrections to the rate of the decay $b \to c \, \overline
{q}_2 \, q_1$ one thus should consider the absorptive part arising from
all possible unitary cuts of the graphs of three types shown in
fig.1.  (These graphs ignore the `penguin' contribution, arising for
$q_1=q_2$, which is to be discussed separately.) In fact, however, it is
clear that the graphs of the type in fig.1c, i.e. with gluon exchange
between the $bc$ line and the $\overline {q}_2 \, q_1$ loop, are vanishing
because of the color trace over the loop (a gluon can not interact through
the loop with two colorless W bosons)\footnote {In particular, this explains
why there is no first-order correction proportional to $\alpha_s  \ln
(m_W/m_b)$: The graphs of the types in fig.1a (gluon exchange on the $bc$
line) and in fig.1b (gluon exchange within the loop) contain renormalization
of the $V-A$ currents, which is not logarithmic.}. The graphs of the type in
fig.1a do not involve the interaction of the quarks $q_1$ and $q_2$ in the
loop with gluons. Therefore the correction arising from these graphs can be
adapted$^{\cite{cm}}$ from the old calculations$^{\cite{muqed}}$ of QED
corrections to the muon decay (see also \cite{corbo,nir}). As to the effects
of the gluon exchange within the colorless $\overline {q}_2 \, q_1$ loop,
these were discussed thus far for both quarks being massless, in which case
the correction reduces to the familiar factor $\alpha_s/\pi$. This limit is
justified for the case of the decay $b \to  c\,{\overline u}\, d$, but, as
will be shown, is misleading for the decay $b \to  c\,{\overline c}\, s$,
where the charmed quark in the loop has mass, which is not small in
comparison with $m_b\,$.

Thus in the first order in $\alpha_s$ the rate of the decay
$b \to c \, \overline c \, s$ can be written as
\beq
\Gamma_{c \overline c s}=\Gamma_{c \overline c s}^{(bare)} \left ( 1+
{\alpha_s \over \pi} \, \bigl [\delta_{bc} + \delta_{\overline c
s}+\delta_{penguin}\bigr ] \right )~,
\label{genexp}
\eeq
where $\Gamma_{c \overline c s}^{(bare)}$ is the rate without any QCD
corrections, $\delta_{bc}$ arises from gluon exchange on the $bc$ line,
$\delta_{\overline c s}$ is due to gluon interactions within the $\overline
c s$ loop, and $\delta_{penguin}$ is due to effects of the penguin type. It
is the goal of the present note to calculate the correction factor
$\delta_{\overline c s}$.

Starting with the relevant term in the weak Lagrangian of the form
\beq
L_{\rm int}=2 \sqrt{2} \, G_F \, V_{cb} V_{q_2 q_1} \bigl ( \overline c_L \,
\gamma_\mu \, b_L \bigr ) \bigl ( \overline q_{1\, L} \, \gamma_\mu q_{2 \,
L} \bigr )
\label{lagr}
\eeq
and parametrizing the spectral density of the current $j_\mu=
\bigl ( \overline q_{1\, L} \, \gamma_\mu q_{2 \,
L} \bigr )$ as
\beq
\sum_n \langle 0 | j^\dagger_\nu (-q) | n \rangle \langle n | j_\mu (q) | 0
\rangle = -{3 \over 8\pi}\,A(q^2) \left ( q^2\,g_{\mu \nu}-q_\mu q_\nu
\right ) + {3 \over 8\pi} B(q^2)\, q_\mu q_\nu~,
\label{sd}
\eeq
one can
write the total decay rate of $b \to c \, \overline {q}_2 \, q_1$ as
\begin{eqnarray}
&& \Gamma_{c \, \overline
{q}_2 \, q_1}=
6 \Gamma_0 \, m_b^{-8}\, \int_{(m_1+m_2)^2}^{(m_b-m_c)^2}
\left \{ A(q^2)\, q^2 \, (m_b^2+m_c^2-q^2) + \right . \nonumber \\
&& \left. {1 \over 2}\,(A(q^2)+B(q^2))\,
 \bigl [ (m_b^2-m_c^2)^2- q^2\, (m_b^2+m_c^2) \bigr ] \right \} \,
\sqrt{\lambda(m_b^2,\,q^2,\,m_c^2)}\, dq^2 ~,
\label{psi}
\end{eqnarray}
where $\lambda(x,\, y,\,z)=x^2+y^2+z^2-2xy-2xz-2yz$ is the standard
kinematical function, and
\beq
\Gamma_0={{3 G_F^2 |V_{cb} V_{q_2 q_1}|^2 m_b^5} \over {192 \pi^3}}
\label{gamma0}
\eeq
is the lowest-order parton decay rate with massless quarks in the final
state.

It should be emphasized that the equation (\ref{psi}) is applicable to
calculation of only the effects associated with the gluon exchanges within
the $\overline q_2 q_1$ loop, i.e. the ones, which are discussed in this
note, and does not include the effects of gluon exchange on the $bc$ line
(like the one shown in fig.1a) or the effects of gluon exchange between the
loop and the $bc$ line, which arise starting from order $\alpha_s^2$.
In the absence of QCD radiative effects the form factors $A$ and $B$ are
readily calculable. In the case when $q_2=c$ and $q_1=s$, so that $m_1=0$
and  $m_2=m_c$, one finds
\begin{eqnarray}
&& A_0={2 \over 3}\,\left ( {{q^2-m_c^2}\over {q^2}} \right
)^2 \, \left ( 1+ {m_c^2 \over {2\,q^2}} \right ) \nonumber \\
&& B_0={m_c^2 \over q^2} \, \left ( {{q^2-m_c^2} \over q^2} \right )^2~.
\label{a0b0}
\end{eqnarray}
Using these expressions in eq.(\ref{psi}) and integrating over $q^2$ gives
the well known result for the `bare' rate of the decay
$b \to c\,{\overline c}\, s$ in eq.(\ref{genexp}):
\beq
\Gamma_{c \overline c s}^{(bare)}=\Gamma_0 \, I
\left ({m_c \over m_b},\, {m_c \over
m_b} \right )~,
\label{gbare}
\eeq
where
\beq
I(x,\,x)=\sqrt{1-4x^2}\, (1-14x^2-2x^4-12x^6)+24 x^4 \,(1-x^4)\, \ln \left (
{{1+\sqrt{1-4x^2}} \over {1-\sqrt{1-4x^2}}} \right )~.
\label{ixx}
\eeq

The QCD radiative corrections due to interactions within the $\overline c s$
loop are expressed through the radiative corrections to the form factors
$A$ and $B$ in the spectral density in eq.(\ref{sd}). The calculation of
the $O(\alpha_s)$ corrections to the spectral density with
unequal masses of quarks has been done in connection with the QCD sum rules
both for the longitudinal form factor $B\,^{\cite{rry1,b,ae}}$ and for
the transversal one $A\, ^{\cite{rry2}}$. For the case, relevant here, where
one of the quarks is massless the result reads as
\begin{eqnarray}
&& A(q^2)=A_0(q^2) \left [ 1 + {4 \over 3} {\alpha_s \over \pi} \left (
f_1 \left ( {q^2 \over m_c^2} \right ) + {{2q^2} \over {2q^2 + m_c^2}}\, f_2
\left ( {q^2 \over m_c^2} \right ) \right ) \right ] \nonumber \\
&& B(q^2)=B_0(q^2) \left [ 1 + {4 \over 3} {\alpha_s \over \pi}
\left ( f_1 \left ( {q^2 \over m_c^2} \right ) -1 \right ) \right ]
\label{ab}
\end{eqnarray}
with
\beq
f_1(z)={13 \over 4} + 2\, {\rm Li} \left ( {1 \over z } \right ) + \ln z \,
\ln {z \over {z-1}} -{3 \over 2} \ln (z-1) + \ln {z \over {z-1}} +{1 \over
z} \ln (z-1) + {1 \over {z-1}} \ln z
\label{f1}
\eeq
and
\beq f_2(z)=-{5
\over 2}-{1 \over z} - {1 \over {z-1}} + \left ( {{z-1} \over z} \right )
\left ( {3 \over 2} + {1 \over 2z} \right ) \ln (z-1) + {z \over {(z-1)^2}}
\ln z~,
\label{f2}
\eeq
where ${\rm Li}(x)=-\int_0^x \ln (1-t) \, dt/t$ is the standard dilogarithm
function\footnote {The correction to the longitudinal form factor $B$
coincides with the correction$^{\cite{rry1}}$ for scalar or pseudoscalar
density up to an additive constant, corresponding to the normalization of
the (pseudo)scalar operator in order $\alpha_s$. This constant is
fixed$^{\cite{b,ae}}$ unambiguously for the longitudinal part of the
vector or axial current. I am thankful to P. Ball for pointing out to me the
papers \cite{b,ae}, where this point is clarified.}.

The integral in eq.(\ref{psi}) with the radiatively corrected values of the
form factors can be easily done numerically, thus giving the value of the
correction factor $\delta_{\overline c s}$ in the equation (\ref{genexp}).
The results of such calculation are shown in fig.2 in terms of the behavior
of $\delta_{\overline c s}$ as a function of $m_c/m_b$. In particular, at a
`reference' point $m_c/m_b=0.3$ one finds $\delta_{\overline c s} =4.46$,
which is significantly larger than the analogous correction factor for the
$b \to  c\,{\overline u}\, d$: $\delta_{\overline u d}=1$.
If one uses the value $\alpha_s=0.2$, then the discussed correction is about
30\%. However, a closer inspection of the integral for the correction in
eq.(\ref{psi}) shows that the integrand has a maximum at $q^2 \approx 2
m_c^2 \approx 0.2 m_b^2$ for the discussed here range of values of
$m_c/m_b$.  Therefore the appropriate value of $\alpha_s$ can in fact be
larger. Naturally, a quantitative clarification of this point requires a
higher order calculation\footnote{It has been recently argued$^{\cite{lsw}}$
that the natural normalization scale for $\alpha_s$ in $\delta_{bc}$ is also
quite low.}.  It can be also noted that the enhancement of the contribution
of relatively low values of $q^2$ is due to the logarithmic growth of the
function $f_1(z)$ in eq.(\ref{f1}) in the threshold region $z \to 1$, which
is a consequence of the `hybrid' anomalous dimension$^{\cite{sv2}}$ of the
current $(\overline c \Gamma s)$.

We therefore conclude, that the decay $b \to
c\,{\overline c}\, s$ is enhanced by about 30\% or more by the correction
proportional to $\delta_{\overline c s}$. To assess the resulting fraction
of this decay in the total decay rate one should also take into account the
corrections with $\delta_{bc}$ and $\delta_{penguin}$ and measure the result
against, say, the semileptonic mode $b \to c\, l\, \nu$ with $l=e$ or
$l=\mu$, whose rate contains only the QCD correction associated with the
$bc$ line i.e. with $\delta^{(l)}_{bc}$. The penguin effect is negative and
is about 3\% - 5\% in magnitude$^{\cite{ap}}$. The term $\delta_{bc}$ is
also negative, but due to the charmed quark mass its magnitude is somewhat
smaller than that of the negative $\delta^{(l)}_{bc}$. In effect the term
with $\delta_{penguin}$ approximately cancels against the difference
$\delta_{bc}- \delta^{(l)}_{bc}$ in the ratio $\Gamma_{c \overline c
s}/\Gamma_{cl\nu}$ and the net $O(\alpha_s)$ correction is dominated by the
large $\delta_{\overline c s}$. Therefore it is quite possible that a
sizeable part of the existing theoretical deficiency of nonleptonic decays
of $B$ can be eliminated by a 30\% or larger enhancement of the decay
$b \to c\,{\overline c}\, s$.

In order to completely quantify the issue of QCD radiative effects in the
$B$ decay rates and to possibly achieve an accuracy of about 5\% in
theoretical predictions for the rate of each inclusive mode a complete
calculation in second order in $\alpha_s$ is needed. Though, no attempt of
such calculation is done in this note, I would like to conclude with a
simple general remark concerning  a calculation of the $O(\alpha_s^2)$
corrections to the inclusive decay rates of the $b$ quark by the unitary
cuts of graphs similar to those in fig.1, albeit in the next order of the
QCD perturbation theory. Also in that order one can split the graphs into
few classes. One class is where the gluons are attached only to the quarks
on the $bc$ line. For the dominant decay $b \to c\, \overline u d$ this
correction cancels in the ratio $\Gamma_{c \overline u d}/\Gamma_{c l \nu}$.
Another is where the gluon corrections are fully contained within the
$\overline q_2 q_1$ loop, which are reduced to the corrections to the form
factors $A$ and $B$ in the spectral density (\ref{sd}).
For the pair of massless quarks $\overline u d$ this can be read off the
corresponding calculation$^{\cite{ds}}$ for $e^+\,e^-$ annihilation into
light hadrons.  Third class is where one gluon is exchanged on the $bc$ line
and the other within the $\overline q_2 q_1$ loop and is thus a product of
the first-order corrections. Finally, because of the color trace over the
loop, the gluon exchange between the $bc$ line and the $\overline q_2 q_1$
loop gives a non-vanishing result only when there are two gluons exchanged,
each starting on the $bc$ line and ending on the $\overline q_2 q_1$ loop,
an example of such graph is shown in fig.3. It is with the latter graphs
that the terms proportional to $\bigl (\alpha_s \ln(m_W/m_b) \bigr )^2$
and to $\alpha_s^2 \ln(m_W/m_b)$
are
associated and which may also contain large non-logarithmic terms.

I am thankful to M. Shifman and A. Vainshtein for discussions of the $B$
decays and to M. Danilov for a discussion of the experimental situation with
the average charm yield in $B$ decays.  This work is supported, in part, by
the DOE grant DE-AC02-83ER40105.

When this paper was finished, a revised version of Ref.\cite{bbbg} appeared,
where a similar estimate of 30\% enhancement of $\Gamma_{c \overline c s}$
was found. I thank P. Ball and V.M. Braun for pointing out to me their
revised estimate.  As discussed in this note, the actual enhancement can in
fact be larger due to a larger value of $\alpha_s$ at the relatively low
relevant invariant mass of the $\overline c s$ pair.

\thicklines
\unitlength=1.00mm
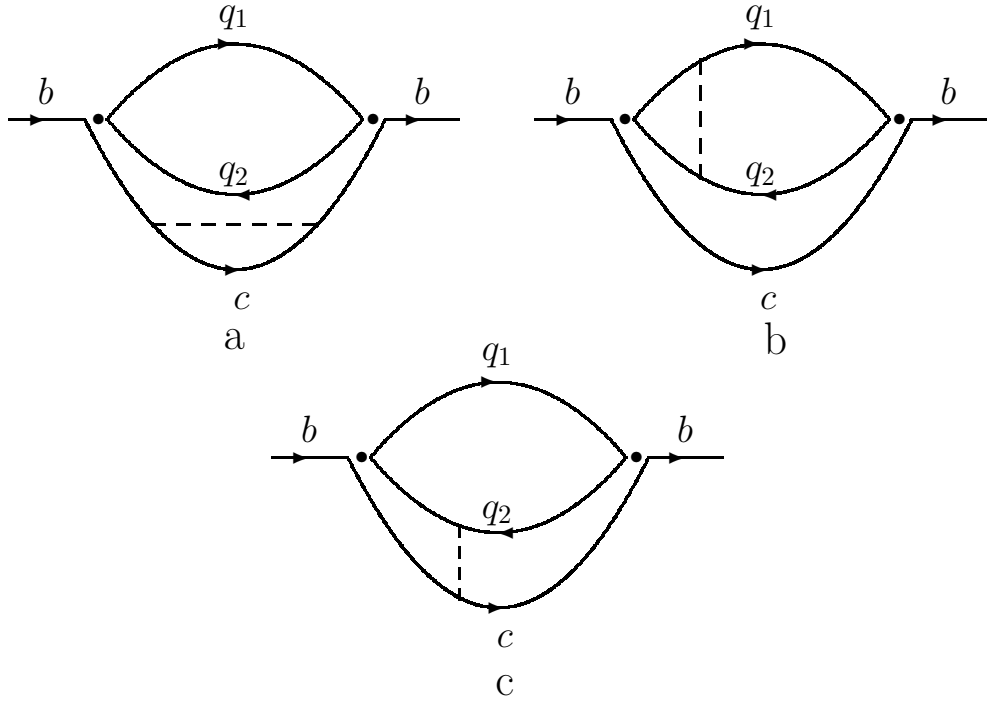
\begin{figure}
\begin{picture}(140.00,100.00)
\put(10.00,80.00){\vector(1,0){5.00}}
\put(15.00,80.00){\line(1,0){5.00}}
\put(60.00,80.00){\vector(1,0){5.00}}
\put(65.00,80.00){\line(1,0){5.00}}
\bezier{356}(20.00,80.00)(40.00,40.00)(60.00,80.00)
\bezier{208}(23.00,80.00)(40.00,60.00)(57.00,80.00)
\bezier{208}(23.00,80.00)(40.00,100.00)(57.00,80.00)
\put(39.00,90.00){\vector(1,0){1.00}}
\put(41.00,70.00){\vector(-1,0){1.00}}
\put(39.00,60.00){\vector(1,0){2.00}}
\put(15.00,82.00){\makebox(0,0)[cb]{\large $b$}}
\put(65.00,82.00){\makebox(0,0)[cb]{\large $b$}}
\put(41.00,57.00){\makebox(0,0)[ct]{\large $c$}}
\put(40.00,93.00){\makebox(0,0)[cb]{\large $q_1$}}
\put(40.00,72.00){\makebox(0,0)[cb]{\large $q_2$}}
\put(22.00,80.00){\circle*{1.50}}
\put(58.50,80.00){\circle*{1.50}}
\put(80.00,80.00){\vector(1,0){5.00}}
\put(85.00,80.00){\line(1,0){5.00}}
\put(130.00,80.00){\vector(1,0){5.00}}
\put(135.00,80.00){\line(1,0){5.00}}
\bezier{356}(90.00,80.00)(110.00,40.00)(130.00,80.00)
\bezier{208}(93.00,80.00)(110.00,60.00)(127.00,80.00)
\bezier{208}(93.00,80.00)(110.00,100.00)(127.00,80.00)
\put(109.00,90.00){\vector(1,0){1.00}}
\put(111.00,70.00){\vector(-1,0){1.00}}
\put(109.00,60.00){\vector(1,0){2.00}}
\put(85.00,82.00){\makebox(0,0)[cb]{\large $b$}}
\put(135.00,82.00){\makebox(0,0)[cb]{\large $b$}}
\put(111.00,57.00){\makebox(0,0)[ct]{\large $c$}}
\put(110.00,93.00){\makebox(0,0)[cb]{\large $q_1$}}
\put(110.00,72.00){\makebox(0,0)[cb]{\large $q_2$}}
\put(92.00,80.00){\circle*{1.50}}
\put(128.50,80.00){\circle*{1.50}}
\put(45.00,35.00){\vector(1,0){5.00}}
\put(50.00,35.00){\line(1,0){5.00}}
\put(95.00,35.00){\vector(1,0){5.00}}
\put(100.00,35.00){\line(1,0){5.00}}
\bezier{356}(55.00,35.00)(75.00,-5.00)(95.00,35.00)
\bezier{208}(58.00,35.00)(75.00,15.00)(92.00,35.00)
\bezier{208}(58.00,35.00)(75.00,55.00)(92.00,35.00)
\put(74.00,45.00){\vector(1,0){1.00}}
\put(76.00,25.00){\vector(-1,0){1.00}}
\put(74.00,15.00){\vector(1,0){2.00}}
\put(50.00,37.00){\makebox(0,0)[cb]{\large $b$}}
\put(100.00,37.00){\makebox(0,0)[cb]{\large $b$}}
\put(76.00,12.00){\makebox(0,0)[ct]{\large $c$}}
\put(75.00,48.00){\makebox(0,0)[cb]{\large $q_1$}}
\put(75.00,27.00){\makebox(0,0)[cb]{\large $q_2$}}
\put(57.00,35.00){\circle*{1.50}}
\put(93.50,35.00){\circle*{1.50}}
\multiput(102.00,72.00)(0,3.50){5}{\line(0,1){2.00}}
\multiput(29.00,66.00)(3.35,0){7}{\line(1,0){1.90}}
\multiput(70.00,16.00)(0,2.80){4}{\line(0,1){1.60}}
\put(40.00,52.00){\makebox(0,0)[ct]{{\Large a}}}
\put(112.00,53.00){\makebox(0,0)[ct]{{\Large b}}}
\put(76.00,6.00){\makebox(0,0)[ct]{{\Large c}}}
\end{picture}
\caption{Three types of graphs, whose unitary cuts describe the first QCD
radiative corrections to the inclusive decay rate $b \to c \overline q_2
q_1$. The small filled circles represent the $W$ boson propagators and the
dashed lines correspond to gluons. The gluon vertices can be anywhere on the
$bc$ line (a), quark lines in the loop (b), or one vertex anywhere on the
$bc$ line and the other vertex on either line in the loop (c).}
\end{figure}

\begin{figure}
\setlength{\unitlength}{0.240900pt}
\ifx\plotpoint\undefined\newsavebox{\plotpoint}\fi
\sbox{\plotpoint}{\rule[-0.200pt]{0.400pt}{0.400pt}}%
\begin{picture}(1500,900)(0,0)
\font\gnuplot=cmr10 at 10pt
\gnuplot
\sbox{\plotpoint}{\rule[-0.200pt]{0.400pt}{0.400pt}}%
\put(220.0,113.0){\rule[-0.200pt]{4.818pt}{0.400pt}}
\put(198,113){\makebox(0,0)[r]{3.5}}
\put(1416.0,113.0){\rule[-0.200pt]{4.818pt}{0.400pt}}
\put(220.0,266.0){\rule[-0.200pt]{4.818pt}{0.400pt}}
\put(198,266){\makebox(0,0)[r]{4}}
\put(1416.0,266.0){\rule[-0.200pt]{4.818pt}{0.400pt}}
\put(220.0,419.0){\rule[-0.200pt]{4.818pt}{0.400pt}}
\put(198,419){\makebox(0,0)[r]{4.5}}
\put(1416.0,419.0){\rule[-0.200pt]{4.818pt}{0.400pt}}
\put(220.0,571.0){\rule[-0.200pt]{4.818pt}{0.400pt}}
\put(198,571){\makebox(0,0)[r]{5}}
\put(1416.0,571.0){\rule[-0.200pt]{4.818pt}{0.400pt}}
\put(220.0,724.0){\rule[-0.200pt]{4.818pt}{0.400pt}}
\put(198,724){\makebox(0,0)[r]{5.5}}
\put(1416.0,724.0){\rule[-0.200pt]{4.818pt}{0.400pt}}
\put(220.0,877.0){\rule[-0.200pt]{4.818pt}{0.400pt}}
\put(198,877){\makebox(0,0)[r]{6}}
\put(1416.0,877.0){\rule[-0.200pt]{4.818pt}{0.400pt}}
\put(220.0,113.0){\rule[-0.200pt]{0.400pt}{4.818pt}}
\put(220,68){\makebox(0,0){0.25}}
\put(220.0,857.0){\rule[-0.200pt]{0.400pt}{4.818pt}}
\put(342.0,113.0){\rule[-0.200pt]{0.400pt}{4.818pt}}
\put(342,68){\makebox(0,0){0.26}}
\put(342.0,857.0){\rule[-0.200pt]{0.400pt}{4.818pt}}
\put(463.0,113.0){\rule[-0.200pt]{0.400pt}{4.818pt}}
\put(463,68){\makebox(0,0){0.27}}
\put(463.0,857.0){\rule[-0.200pt]{0.400pt}{4.818pt}}
\put(585.0,113.0){\rule[-0.200pt]{0.400pt}{4.818pt}}
\put(585,68){\makebox(0,0){0.28}}
\put(585.0,857.0){\rule[-0.200pt]{0.400pt}{4.818pt}}
\put(706.0,113.0){\rule[-0.200pt]{0.400pt}{4.818pt}}
\put(706,68){\makebox(0,0){0.29}}
\put(706.0,857.0){\rule[-0.200pt]{0.400pt}{4.818pt}}
\put(828.0,113.0){\rule[-0.200pt]{0.400pt}{4.818pt}}
\put(828,68){\makebox(0,0){0.3}}
\put(828.0,857.0){\rule[-0.200pt]{0.400pt}{4.818pt}}
\put(950.0,113.0){\rule[-0.200pt]{0.400pt}{4.818pt}}
\put(950,68){\makebox(0,0){0.31}}
\put(950.0,857.0){\rule[-0.200pt]{0.400pt}{4.818pt}}
\put(1071.0,113.0){\rule[-0.200pt]{0.400pt}{4.818pt}}
\put(1071,68){\makebox(0,0){0.32}}
\put(1071.0,857.0){\rule[-0.200pt]{0.400pt}{4.818pt}}
\put(1193.0,113.0){\rule[-0.200pt]{0.400pt}{4.818pt}}
\put(1193,68){\makebox(0,0){0.33}}
\put(1193.0,857.0){\rule[-0.200pt]{0.400pt}{4.818pt}}
\put(1314.0,113.0){\rule[-0.200pt]{0.400pt}{4.818pt}}
\put(1314,68){\makebox(0,0){0.34}}
\put(1314.0,857.0){\rule[-0.200pt]{0.400pt}{4.818pt}}
\put(1436.0,113.0){\rule[-0.200pt]{0.400pt}{4.818pt}}
\put(1436,68){\makebox(0,0){0.35}}
\put(1436.0,857.0){\rule[-0.200pt]{0.400pt}{4.818pt}}
\put(220.0,113.0){\rule[-0.200pt]{292.934pt}{0.400pt}}
\put(1436.0,113.0){\rule[-0.200pt]{0.400pt}{184.048pt}}
\put(220.0,877.0){\rule[-0.200pt]{292.934pt}{0.400pt}}
\put(95,495){\makebox(0,0){{\large $\delta_{\overline c s}$}}}
\put(828,0){\makebox(0,0){{\large $m_c/m_b$}}}
\put(220.0,113.0){\rule[-0.200pt]{0.400pt}{184.048pt}}
\put(220,132){\usebox{\plotpoint}}
\multiput(220.00,132.59)(1.267,0.477){7}{\rule{1.060pt}{0.115pt}}
\multiput(220.00,131.17)(9.800,5.000){2}{\rule{0.530pt}{0.400pt}}
\multiput(232.00,137.59)(1.378,0.477){7}{\rule{1.140pt}{0.115pt}}
\multiput(232.00,136.17)(10.634,5.000){2}{\rule{0.570pt}{0.400pt}}
\multiput(245.00,142.59)(1.267,0.477){7}{\rule{1.060pt}{0.115pt}}
\multiput(245.00,141.17)(9.800,5.000){2}{\rule{0.530pt}{0.400pt}}
\multiput(257.00,147.59)(1.267,0.477){7}{\rule{1.060pt}{0.115pt}}
\multiput(257.00,146.17)(9.800,5.000){2}{\rule{0.530pt}{0.400pt}}
\multiput(269.00,152.59)(1.267,0.477){7}{\rule{1.060pt}{0.115pt}}
\multiput(269.00,151.17)(9.800,5.000){2}{\rule{0.530pt}{0.400pt}}
\multiput(281.00,157.59)(1.123,0.482){9}{\rule{0.967pt}{0.116pt}}
\multiput(281.00,156.17)(10.994,6.000){2}{\rule{0.483pt}{0.400pt}}
\multiput(294.00,163.59)(1.267,0.477){7}{\rule{1.060pt}{0.115pt}}
\multiput(294.00,162.17)(9.800,5.000){2}{\rule{0.530pt}{0.400pt}}
\multiput(306.00,168.59)(1.267,0.477){7}{\rule{1.060pt}{0.115pt}}
\multiput(306.00,167.17)(9.800,5.000){2}{\rule{0.530pt}{0.400pt}}
\multiput(318.00,173.59)(1.378,0.477){7}{\rule{1.140pt}{0.115pt}}
\multiput(318.00,172.17)(10.634,5.000){2}{\rule{0.570pt}{0.400pt}}
\multiput(331.00,178.59)(1.267,0.477){7}{\rule{1.060pt}{0.115pt}}
\multiput(331.00,177.17)(9.800,5.000){2}{\rule{0.530pt}{0.400pt}}
\multiput(343.00,183.59)(1.033,0.482){9}{\rule{0.900pt}{0.116pt}}
\multiput(343.00,182.17)(10.132,6.000){2}{\rule{0.450pt}{0.400pt}}
\multiput(355.00,189.59)(1.267,0.477){7}{\rule{1.060pt}{0.115pt}}
\multiput(355.00,188.17)(9.800,5.000){2}{\rule{0.530pt}{0.400pt}}
\multiput(367.00,194.59)(1.378,0.477){7}{\rule{1.140pt}{0.115pt}}
\multiput(367.00,193.17)(10.634,5.000){2}{\rule{0.570pt}{0.400pt}}
\multiput(380.00,199.59)(1.033,0.482){9}{\rule{0.900pt}{0.116pt}}
\multiput(380.00,198.17)(10.132,6.000){2}{\rule{0.450pt}{0.400pt}}
\multiput(392.00,205.59)(1.267,0.477){7}{\rule{1.060pt}{0.115pt}}
\multiput(392.00,204.17)(9.800,5.000){2}{\rule{0.530pt}{0.400pt}}
\multiput(404.00,210.59)(1.378,0.477){7}{\rule{1.140pt}{0.115pt}}
\multiput(404.00,209.17)(10.634,5.000){2}{\rule{0.570pt}{0.400pt}}
\multiput(417.00,215.59)(1.033,0.482){9}{\rule{0.900pt}{0.116pt}}
\multiput(417.00,214.17)(10.132,6.000){2}{\rule{0.450pt}{0.400pt}}
\multiput(429.00,221.59)(1.267,0.477){7}{\rule{1.060pt}{0.115pt}}
\multiput(429.00,220.17)(9.800,5.000){2}{\rule{0.530pt}{0.400pt}}
\multiput(441.00,226.59)(1.033,0.482){9}{\rule{0.900pt}{0.116pt}}
\multiput(441.00,225.17)(10.132,6.000){2}{\rule{0.450pt}{0.400pt}}
\multiput(453.00,232.59)(1.378,0.477){7}{\rule{1.140pt}{0.115pt}}
\multiput(453.00,231.17)(10.634,5.000){2}{\rule{0.570pt}{0.400pt}}
\multiput(466.00,237.59)(1.267,0.477){7}{\rule{1.060pt}{0.115pt}}
\multiput(466.00,236.17)(9.800,5.000){2}{\rule{0.530pt}{0.400pt}}
\multiput(478.00,242.59)(1.033,0.482){9}{\rule{0.900pt}{0.116pt}}
\multiput(478.00,241.17)(10.132,6.000){2}{\rule{0.450pt}{0.400pt}}
\multiput(490.00,248.59)(1.123,0.482){9}{\rule{0.967pt}{0.116pt}}
\multiput(490.00,247.17)(10.994,6.000){2}{\rule{0.483pt}{0.400pt}}
\multiput(503.00,254.59)(1.267,0.477){7}{\rule{1.060pt}{0.115pt}}
\multiput(503.00,253.17)(9.800,5.000){2}{\rule{0.530pt}{0.400pt}}
\multiput(515.00,259.59)(1.033,0.482){9}{\rule{0.900pt}{0.116pt}}
\multiput(515.00,258.17)(10.132,6.000){2}{\rule{0.450pt}{0.400pt}}
\multiput(527.00,265.59)(1.267,0.477){7}{\rule{1.060pt}{0.115pt}}
\multiput(527.00,264.17)(9.800,5.000){2}{\rule{0.530pt}{0.400pt}}
\multiput(539.00,270.59)(1.123,0.482){9}{\rule{0.967pt}{0.116pt}}
\multiput(539.00,269.17)(10.994,6.000){2}{\rule{0.483pt}{0.400pt}}
\multiput(552.00,276.59)(1.267,0.477){7}{\rule{1.060pt}{0.115pt}}
\multiput(552.00,275.17)(9.800,5.000){2}{\rule{0.530pt}{0.400pt}}
\multiput(564.00,281.59)(1.033,0.482){9}{\rule{0.900pt}{0.116pt}}
\multiput(564.00,280.17)(10.132,6.000){2}{\rule{0.450pt}{0.400pt}}
\multiput(576.00,287.59)(1.033,0.482){9}{\rule{0.900pt}{0.116pt}}
\multiput(576.00,286.17)(10.132,6.000){2}{\rule{0.450pt}{0.400pt}}
\multiput(588.00,293.59)(1.378,0.477){7}{\rule{1.140pt}{0.115pt}}
\multiput(588.00,292.17)(10.634,5.000){2}{\rule{0.570pt}{0.400pt}}
\multiput(601.00,298.59)(1.033,0.482){9}{\rule{0.900pt}{0.116pt}}
\multiput(601.00,297.17)(10.132,6.000){2}{\rule{0.450pt}{0.400pt}}
\multiput(613.00,304.59)(1.033,0.482){9}{\rule{0.900pt}{0.116pt}}
\multiput(613.00,303.17)(10.132,6.000){2}{\rule{0.450pt}{0.400pt}}
\multiput(625.00,310.59)(1.123,0.482){9}{\rule{0.967pt}{0.116pt}}
\multiput(625.00,309.17)(10.994,6.000){2}{\rule{0.483pt}{0.400pt}}
\multiput(638.00,316.59)(1.267,0.477){7}{\rule{1.060pt}{0.115pt}}
\multiput(638.00,315.17)(9.800,5.000){2}{\rule{0.530pt}{0.400pt}}
\multiput(650.00,321.59)(1.033,0.482){9}{\rule{0.900pt}{0.116pt}}
\multiput(650.00,320.17)(10.132,6.000){2}{\rule{0.450pt}{0.400pt}}
\multiput(662.00,327.59)(1.033,0.482){9}{\rule{0.900pt}{0.116pt}}
\multiput(662.00,326.17)(10.132,6.000){2}{\rule{0.450pt}{0.400pt}}
\multiput(674.00,333.59)(1.123,0.482){9}{\rule{0.967pt}{0.116pt}}
\multiput(674.00,332.17)(10.994,6.000){2}{\rule{0.483pt}{0.400pt}}
\multiput(687.00,339.59)(1.033,0.482){9}{\rule{0.900pt}{0.116pt}}
\multiput(687.00,338.17)(10.132,6.000){2}{\rule{0.450pt}{0.400pt}}
\multiput(699.00,345.59)(1.033,0.482){9}{\rule{0.900pt}{0.116pt}}
\multiput(699.00,344.17)(10.132,6.000){2}{\rule{0.450pt}{0.400pt}}
\multiput(711.00,351.59)(1.123,0.482){9}{\rule{0.967pt}{0.116pt}}
\multiput(711.00,350.17)(10.994,6.000){2}{\rule{0.483pt}{0.400pt}}
\multiput(724.00,357.59)(1.033,0.482){9}{\rule{0.900pt}{0.116pt}}
\multiput(724.00,356.17)(10.132,6.000){2}{\rule{0.450pt}{0.400pt}}
\multiput(736.00,363.59)(1.033,0.482){9}{\rule{0.900pt}{0.116pt}}
\multiput(736.00,362.17)(10.132,6.000){2}{\rule{0.450pt}{0.400pt}}
\multiput(748.00,369.59)(1.033,0.482){9}{\rule{0.900pt}{0.116pt}}
\multiput(748.00,368.17)(10.132,6.000){2}{\rule{0.450pt}{0.400pt}}
\multiput(760.00,375.59)(1.123,0.482){9}{\rule{0.967pt}{0.116pt}}
\multiput(760.00,374.17)(10.994,6.000){2}{\rule{0.483pt}{0.400pt}}
\multiput(773.00,381.59)(1.033,0.482){9}{\rule{0.900pt}{0.116pt}}
\multiput(773.00,380.17)(10.132,6.000){2}{\rule{0.450pt}{0.400pt}}
\multiput(785.00,387.59)(1.033,0.482){9}{\rule{0.900pt}{0.116pt}}
\multiput(785.00,386.17)(10.132,6.000){2}{\rule{0.450pt}{0.400pt}}
\multiput(797.00,393.59)(1.123,0.482){9}{\rule{0.967pt}{0.116pt}}
\multiput(797.00,392.17)(10.994,6.000){2}{\rule{0.483pt}{0.400pt}}
\multiput(810.00,399.59)(1.033,0.482){9}{\rule{0.900pt}{0.116pt}}
\multiput(810.00,398.17)(10.132,6.000){2}{\rule{0.450pt}{0.400pt}}
\multiput(822.00,405.59)(1.033,0.482){9}{\rule{0.900pt}{0.116pt}}
\multiput(822.00,404.17)(10.132,6.000){2}{\rule{0.450pt}{0.400pt}}
\multiput(834.00,411.59)(1.033,0.482){9}{\rule{0.900pt}{0.116pt}}
\multiput(834.00,410.17)(10.132,6.000){2}{\rule{0.450pt}{0.400pt}}
\multiput(846.00,417.59)(1.123,0.482){9}{\rule{0.967pt}{0.116pt}}
\multiput(846.00,416.17)(10.994,6.000){2}{\rule{0.483pt}{0.400pt}}
\multiput(859.00,423.59)(1.033,0.482){9}{\rule{0.900pt}{0.116pt}}
\multiput(859.00,422.17)(10.132,6.000){2}{\rule{0.450pt}{0.400pt}}
\multiput(871.00,429.59)(0.874,0.485){11}{\rule{0.786pt}{0.117pt}}
\multiput(871.00,428.17)(10.369,7.000){2}{\rule{0.393pt}{0.400pt}}
\multiput(883.00,436.59)(1.123,0.482){9}{\rule{0.967pt}{0.116pt}}
\multiput(883.00,435.17)(10.994,6.000){2}{\rule{0.483pt}{0.400pt}}
\multiput(896.00,442.59)(1.033,0.482){9}{\rule{0.900pt}{0.116pt}}
\multiput(896.00,441.17)(10.132,6.000){2}{\rule{0.450pt}{0.400pt}}
\multiput(908.00,448.59)(0.874,0.485){11}{\rule{0.786pt}{0.117pt}}
\multiput(908.00,447.17)(10.369,7.000){2}{\rule{0.393pt}{0.400pt}}
\multiput(920.00,455.59)(1.033,0.482){9}{\rule{0.900pt}{0.116pt}}
\multiput(920.00,454.17)(10.132,6.000){2}{\rule{0.450pt}{0.400pt}}
\multiput(932.00,461.59)(1.123,0.482){9}{\rule{0.967pt}{0.116pt}}
\multiput(932.00,460.17)(10.994,6.000){2}{\rule{0.483pt}{0.400pt}}
\multiput(945.00,467.59)(0.874,0.485){11}{\rule{0.786pt}{0.117pt}}
\multiput(945.00,466.17)(10.369,7.000){2}{\rule{0.393pt}{0.400pt}}
\multiput(957.00,474.59)(1.033,0.482){9}{\rule{0.900pt}{0.116pt}}
\multiput(957.00,473.17)(10.132,6.000){2}{\rule{0.450pt}{0.400pt}}
\multiput(969.00,480.59)(1.123,0.482){9}{\rule{0.967pt}{0.116pt}}
\multiput(969.00,479.17)(10.994,6.000){2}{\rule{0.483pt}{0.400pt}}
\multiput(982.00,486.59)(0.874,0.485){11}{\rule{0.786pt}{0.117pt}}
\multiput(982.00,485.17)(10.369,7.000){2}{\rule{0.393pt}{0.400pt}}
\multiput(994.00,493.59)(1.033,0.482){9}{\rule{0.900pt}{0.116pt}}
\multiput(994.00,492.17)(10.132,6.000){2}{\rule{0.450pt}{0.400pt}}
\multiput(1006.00,499.59)(0.874,0.485){11}{\rule{0.786pt}{0.117pt}}
\multiput(1006.00,498.17)(10.369,7.000){2}{\rule{0.393pt}{0.400pt}}
\multiput(1018.00,506.59)(1.123,0.482){9}{\rule{0.967pt}{0.116pt}}
\multiput(1018.00,505.17)(10.994,6.000){2}{\rule{0.483pt}{0.400pt}}
\multiput(1031.00,512.59)(0.874,0.485){11}{\rule{0.786pt}{0.117pt}}
\multiput(1031.00,511.17)(10.369,7.000){2}{\rule{0.393pt}{0.400pt}}
\multiput(1043.00,519.59)(0.874,0.485){11}{\rule{0.786pt}{0.117pt}}
\multiput(1043.00,518.17)(10.369,7.000){2}{\rule{0.393pt}{0.400pt}}
\multiput(1055.00,526.59)(1.123,0.482){9}{\rule{0.967pt}{0.116pt}}
\multiput(1055.00,525.17)(10.994,6.000){2}{\rule{0.483pt}{0.400pt}}
\multiput(1068.00,532.59)(0.874,0.485){11}{\rule{0.786pt}{0.117pt}}
\multiput(1068.00,531.17)(10.369,7.000){2}{\rule{0.393pt}{0.400pt}}
\multiput(1080.00,539.59)(0.874,0.485){11}{\rule{0.786pt}{0.117pt}}
\multiput(1080.00,538.17)(10.369,7.000){2}{\rule{0.393pt}{0.400pt}}
\multiput(1092.00,546.59)(1.033,0.482){9}{\rule{0.900pt}{0.116pt}}
\multiput(1092.00,545.17)(10.132,6.000){2}{\rule{0.450pt}{0.400pt}}
\multiput(1104.00,552.59)(0.950,0.485){11}{\rule{0.843pt}{0.117pt}}
\multiput(1104.00,551.17)(11.251,7.000){2}{\rule{0.421pt}{0.400pt}}
\multiput(1117.00,559.59)(0.874,0.485){11}{\rule{0.786pt}{0.117pt}}
\multiput(1117.00,558.17)(10.369,7.000){2}{\rule{0.393pt}{0.400pt}}
\multiput(1129.00,566.59)(0.874,0.485){11}{\rule{0.786pt}{0.117pt}}
\multiput(1129.00,565.17)(10.369,7.000){2}{\rule{0.393pt}{0.400pt}}
\multiput(1141.00,573.59)(0.874,0.485){11}{\rule{0.786pt}{0.117pt}}
\multiput(1141.00,572.17)(10.369,7.000){2}{\rule{0.393pt}{0.400pt}}
\multiput(1153.00,580.59)(1.123,0.482){9}{\rule{0.967pt}{0.116pt}}
\multiput(1153.00,579.17)(10.994,6.000){2}{\rule{0.483pt}{0.400pt}}
\multiput(1166.00,586.59)(0.874,0.485){11}{\rule{0.786pt}{0.117pt}}
\multiput(1166.00,585.17)(10.369,7.000){2}{\rule{0.393pt}{0.400pt}}
\multiput(1178.00,593.59)(0.874,0.485){11}{\rule{0.786pt}{0.117pt}}
\multiput(1178.00,592.17)(10.369,7.000){2}{\rule{0.393pt}{0.400pt}}
\multiput(1190.00,600.59)(0.950,0.485){11}{\rule{0.843pt}{0.117pt}}
\multiput(1190.00,599.17)(11.251,7.000){2}{\rule{0.421pt}{0.400pt}}
\multiput(1203.00,607.59)(0.874,0.485){11}{\rule{0.786pt}{0.117pt}}
\multiput(1203.00,606.17)(10.369,7.000){2}{\rule{0.393pt}{0.400pt}}
\multiput(1215.00,614.59)(0.874,0.485){11}{\rule{0.786pt}{0.117pt}}
\multiput(1215.00,613.17)(10.369,7.000){2}{\rule{0.393pt}{0.400pt}}
\multiput(1227.00,621.59)(0.874,0.485){11}{\rule{0.786pt}{0.117pt}}
\multiput(1227.00,620.17)(10.369,7.000){2}{\rule{0.393pt}{0.400pt}}
\multiput(1239.00,628.59)(0.824,0.488){13}{\rule{0.750pt}{0.117pt}}
\multiput(1239.00,627.17)(11.443,8.000){2}{\rule{0.375pt}{0.400pt}}
\multiput(1252.00,636.59)(0.874,0.485){11}{\rule{0.786pt}{0.117pt}}
\multiput(1252.00,635.17)(10.369,7.000){2}{\rule{0.393pt}{0.400pt}}
\multiput(1264.00,643.59)(0.874,0.485){11}{\rule{0.786pt}{0.117pt}}
\multiput(1264.00,642.17)(10.369,7.000){2}{\rule{0.393pt}{0.400pt}}
\multiput(1276.00,650.59)(0.950,0.485){11}{\rule{0.843pt}{0.117pt}}
\multiput(1276.00,649.17)(11.251,7.000){2}{\rule{0.421pt}{0.400pt}}
\multiput(1289.00,657.59)(0.874,0.485){11}{\rule{0.786pt}{0.117pt}}
\multiput(1289.00,656.17)(10.369,7.000){2}{\rule{0.393pt}{0.400pt}}
\multiput(1301.00,664.59)(0.758,0.488){13}{\rule{0.700pt}{0.117pt}}
\multiput(1301.00,663.17)(10.547,8.000){2}{\rule{0.350pt}{0.400pt}}
\multiput(1313.00,672.59)(0.874,0.485){11}{\rule{0.786pt}{0.117pt}}
\multiput(1313.00,671.17)(10.369,7.000){2}{\rule{0.393pt}{0.400pt}}
\multiput(1325.00,679.59)(0.950,0.485){11}{\rule{0.843pt}{0.117pt}}
\multiput(1325.00,678.17)(11.251,7.000){2}{\rule{0.421pt}{0.400pt}}
\multiput(1338.00,686.59)(0.758,0.488){13}{\rule{0.700pt}{0.117pt}}
\multiput(1338.00,685.17)(10.547,8.000){2}{\rule{0.350pt}{0.400pt}}
\multiput(1350.00,694.59)(0.874,0.485){11}{\rule{0.786pt}{0.117pt}}
\multiput(1350.00,693.17)(10.369,7.000){2}{\rule{0.393pt}{0.400pt}}
\multiput(1362.00,701.59)(0.824,0.488){13}{\rule{0.750pt}{0.117pt}}
\multiput(1362.00,700.17)(11.443,8.000){2}{\rule{0.375pt}{0.400pt}}
\multiput(1375.00,709.59)(0.874,0.485){11}{\rule{0.786pt}{0.117pt}}
\multiput(1375.00,708.17)(10.369,7.000){2}{\rule{0.393pt}{0.400pt}}
\multiput(1387.00,716.59)(0.758,0.488){13}{\rule{0.700pt}{0.117pt}}
\multiput(1387.00,715.17)(10.547,8.000){2}{\rule{0.350pt}{0.400pt}}
\multiput(1399.00,724.59)(0.874,0.485){11}{\rule{0.786pt}{0.117pt}}
\multiput(1399.00,723.17)(10.369,7.000){2}{\rule{0.393pt}{0.400pt}}
\multiput(1411.00,731.59)(0.824,0.488){13}{\rule{0.750pt}{0.117pt}}
\multiput(1411.00,730.17)(11.443,8.000){2}{\rule{0.375pt}{0.400pt}}
\multiput(1424.00,739.59)(0.758,0.488){13}{\rule{0.700pt}{0.117pt}}
\multiput(1424.00,738.17)(10.547,8.000){2}{\rule{0.350pt}{0.400pt}}
\end{picture}

\caption{The correction factor $\delta_{\overline c s}$ in
eq.(1), arising from the gluon exchange within the $\overline c
s$ loop, vs. the mass ratio $m_c/m_b$. The range of $m_c/m_b$ shown well
covers the ratio of the actual quark masses with the existing uncertainty.}
\end{figure}
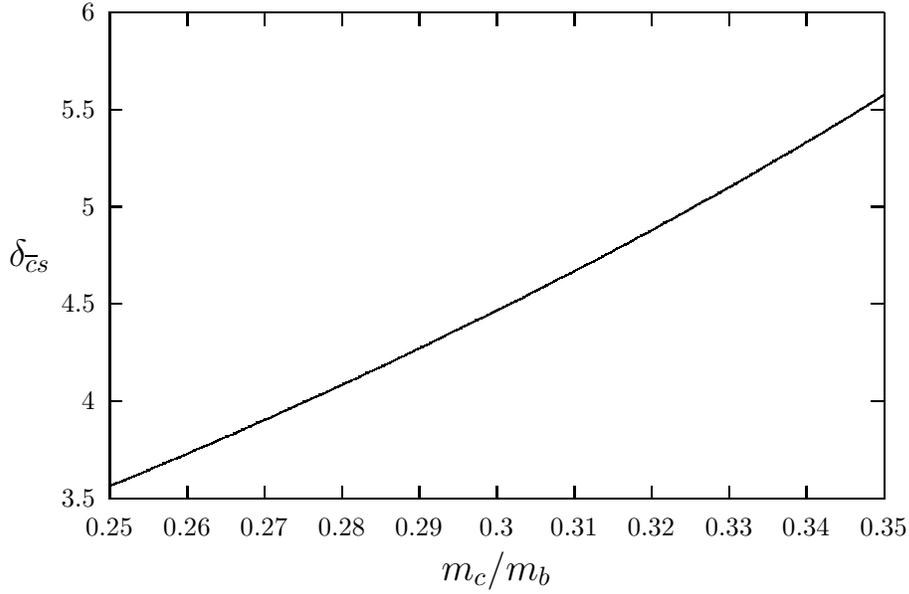

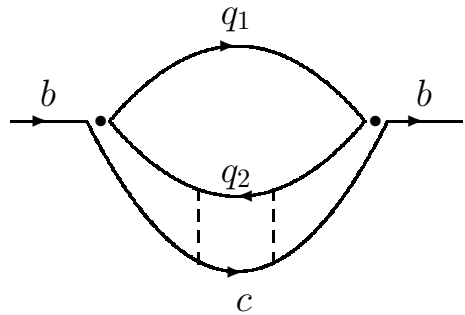
\begin{figure}
\unitlength 1.00mm
\begin{picture}(140.00,55.00)
\put(45.00,35.00){\vector(1,0){5.00}}
\put(50.00,35.00){\line(1,0){5.00}}
\put(95.00,35.00){\vector(1,0){5.00}}
\put(100.00,35.00){\line(1,0){5.00}}
\bezier{356}(55.00,35.00)(75.00,-5.00)(95.00,35.00)
\bezier{208}(58.00,35.00)(75.00,15.00)(92.00,35.00)
\bezier{208}(58.00,35.00)(75.00,55.00)(92.00,35.00)
\put(74.00,45.00){\vector(1,0){1.00}}
\put(76.00,25.00){\vector(-1,0){1.00}}
\put(74.00,15.00){\vector(1,0){2.00}}
\put(50.00,37.00){\makebox(0,0)[cb]{\large $b$}}
\put(100.00,37.00){\makebox(0,0)[cb]{\large $b$}}
\put(76.00,12.00){\makebox(0,0)[ct]{\large $c$}}
\put(75.00,48.00){\makebox(0,0)[cb]{\large $q_1$}}
\put(75.00,27.00){\makebox(0,0)[cb]{\large $q_2$}}
\put(57.00,35.00){\circle*{1.50}}
\put(93.50,35.00){\circle*{1.50}}
\multiput(80.00,16.00)(0,2.80){4}{\line(0,1){1.60}}
\multiput(70.00,16.00)(0,2.80){4}{\line(0,1){1.60}}
\end{picture}
\caption{The only type of graphs in the second order in $\alpha_s$ with
gluon exchange between the $bc$ line and the $\overline q_2 q_1$ loop, which
gives contribution to the inclusive decay rate. Each of the two gluons
should start anywhere on the $bc$ line and end on either of the quark lines
in the loop.}
\end{figure}

\end{document}